# Superhard **Ion** $C_5$ and derived carbon nitrides: $C_4N$ and $C_2N_2$. Crystal chemistry and first principles DFT studies.

Samir F. Matar

Lebanese German University (LGU), Sahel Alma, P.O. Box 206 Jounieh, Lebanon

https://orcid.org/0000-0001-5419-358X

Email: s.matar@lgu.edu.lb & abouliess@gmail.com

*Abstract*

*Super-hard $C_5$ with **Ion** topology (**Ion**: Lonsdaleite hexagonal diamond) and characterized by the presence of $sp^3$ and $sp^2$ –like carbon sites is devised from crystal chemistry and used as template matrix structure for identifying original carbonitrides $C_4N$ and $C_2N_2$ with **Ion** topology except for the equiatomic belonging to a new topology (3,4**L**147). The steric effect of $N(2s^2)$ lone pair is highlighted in $C_2N_2$ in inducing an original structure of largely separated two-layered stacking of tetrahedra. The investigations based on crystal chemistry were backed by computations within the quantum density functional theory DFT. All systems were found cohesive and both mechanically (elastic constants) and dynamically (phonons band structures) stable. Super hardness characterizes the carbon allotrope $C_5$ and the nitrides $C_4N$ and $C_2N_2$. Metallic-like conductivities and insulating characters were identified.*

**Keywords**

Carbon; carbon nitrides; DFT; Phonons; Hardness; topology.

**Introduction**

Carbon allotropes, especially those that approach diamond's physical properties (mechanical, electronic, thermodynamic, ….) benefit from continuous interest within the scientific community (cf. [1] for a mini review). Properties as hardness arising from the three-dimensional 3D arrangements of *C4* tetrahedra are possessed by both the prevailing cubic form as well as in the less common hexagonal allotrope 'Lonsdaleite'. Topology-wise [2], the cubic and the hexagonal allotropes are the aristo-types labeled **dia** and **lon**, respectively. Other families of carbon allotropes follow these nomenclatures. Carbon insertions with other carbon hybridizations like $sp^2$ and $sp^1$ into diamond's $C(sp^3)$ tetrahedral lattice led to novel allotropes with original physical properties [3].

Expanding to carbon neighbors B and N, boron nitride BN as well as carbon nitrides such as $C_3N_4$ are superhard, albeit with lower hardness than diamond. They were early synthesized to replace expensive diamond, unstable at elevated temperatures, in mechanical applications (cf. review paper [4]). The research efforts to identify new chemical systems call for structure prediction programs like CALYPSO [5] and USPEX [6] and more recently machine learning crystallography code *CrystalMELA* [7]. Nevertheless, besides the opportunity offered by modern machine learned codes, novel structures with original properties as hardness and electronic structures can be identified from crystal engineering rationale as presented herein. In all cases such predictions need validation through quantitative studies of the energies and the derived physical properties with the help of first principles calculations. Throughout the years the well-established quantum mechanics framework of the density functional theory (DFT) [8,9] has been shown most efficient.

In the present paper, we further develop the topic of novel compounds based on light elements within the B-C-N diagram [4] by focusing on original carbon nitrides. Firstly, we propose a novel cohesive carbon allotrope; hexagonal $C_5$ through crystal chemistry schematics going from two dimensional 2D carbon to three-dimensional 3D carbon through rationalized C insertion; the 3D allotrope $C_5$ was found to belong alike Lonsdaleite to **lon** topology (3D **lon**-$C_5$). A similar protocol was used to derive 3D **lon**-$C_6$ from 2D carbon, for the sake of comparing the different physical properties between the two allotropes, mainly the mechanical ones in view of the different local structures.

$C_5$ was then used as a template for selective substitutions of carbon for nitrogen leading to the nitride $C_4N$ with **Ion** topology on one hand, and to the equiatomic $C_2N_2$ on the other hand with characteristic layered structure of tetrahedral stacking, found to present a new topology (3,4**L**147). Particularly, the steric effect of $N(2s^2)$ lone pair in tailoring $C_2N_2$ into layered-like structure while keeping 3D dimensionality and the elevated hardness are closely examined. The investigations will show that all novel chemical systems (allotropes and phases) are found cohesive and stable both mechanically as well as dynamically and characterized by elevated Vickers hardness with metallic-like behavior for $C_6$, $C_5$ allotropes and $C_4N$ on one hand, and insulating behavior for the equiatomic $C_2N_2$, on the other hand.

## 1- Computational framework

The identifications of the ground state structures corresponding to the energy minima and the subsequent prediction of their mechanical and dynamical properties were carried out by DFT-based calculations. The Vienna Ab initio Simulation Package (VASP) code [10,11] and the projector augmented wave (PAW) method [11,12] for the atomic potentials were used. DFT exchange-correlation (XC) effects were considered using the generalized gradient functional approximation (GGA) [13]. The relaxation of atoms onto the ground state structures was performed with the conjugate gradient algorithm according to Press *et al*. [14]. The Blöchl tetrahedron method [15] with corrections according to the Methfessel and Paxton scheme [16] was used for geometry optimization and energy calculations, respectively. Brillouin-zone (BZ) integrals were approximated by a special **k**-point sampling according to Monkhorst and Pack [17]. Structural parameters were optimized until atomic forces were below 0.02 eV/Å and all stress components were < 0.003 eV/Å$^3$. The calculations were converged at an energy cutoff of 400 eV for the plane-wave basis set in terms of the **k**-point integration in the reciprocal space from $k_x(6) \times k_y(6) \times k_z(6)$ up to $k_x(12) \times k_y(12) \times k_z(12)$ to obtain a final convergence and relaxation to zero strains for the original stoichiometries presented in this work. In the post-processing of the ground state electronic structures, the charge density projections were operated on the lattice sites.

The mechanical stability criteria were obtained from the calculations of the elastic constants. The treatment of the results was done thanks to ELATE [18] online tool devoted to the analysis of the elastic tensors. The program provides the bulk B and shear G modules along different averaging methods; Voigt's method was used herein [19]. For the calculation of the Vickers hardness a semi-empirical model was used [20]. The dynamic stabilities were confirmed from the phonon positive magnitudes. The corresponding phonon band structures were obtained from a high resolution of the cubic and hexagonal Brillouin Zones (BZ) according to Togo *et al*. [21]. The electronic band structures were obtained using the all-electron DFT based ASW method [22] and the GGA XC functional [13]. The VESTA (Visualization for Electronic and Structural Analysis) program [23] was used to visualize the crystal structures and charge densities.

2- **Crystal chemistry rationale**

*2a- 2D/3D Carbon systems*

Lonsdaleite (space group $P6_3/mmc$ No. 194) is characterized by a single 4-fold atomic position for carbon at 1/3, 2/3, $z$ (Fig. 1a), with $z = 0.06275$. Such a low magnitude of the coordinate along the hexagonal vertical axis is related to the off-plane carbon atomic arrangement. Indeed, by changing z to 0 a graphitic structure is obtained (Fig. 1b) with non-eclipsed (staggered) planes. The space group is kept the same (cf. Table 1). Then, the puckering of the graphitic layers is a key factor in going from 2D to 3D structures. Such transformation: Graphite $\rightarrow$ Diamond is usually induced by high temperature/high pressure (HT/HP) conditions found in the Earth mantle near volcanoes or between tectonic plates with very slow kinetics. Inducing the 2D $\rightarrow$ 3D carbon transformation can be modeled by inserting carbon (white spheres) between layers as shown in Fig. 1c. Geometry relaxation leads anew to a 3D structure with $C_6$ stoichiometry (Fig. 1d). A relevant feature is the *zigzag* (or shifted) alignment of the *C3* entities (brown-white-brown spheres) within the cell that will be opposed to the other $C_5$ allotrope showing eclipsed-like arrangement (*vide infra*). The crystal data are presented in Table 1. Focusing on the ground state energies, the cohesive energies were obtained by subtracting the atomic energy of isolated carbon in a box (-6.6 eV) from the total energy, then by averaging as per atom – the same process is done for the other compounds in this paper. The cohesive energies of the three allotropes belonging to $P6_3/mmc$, No. 194, the planar 2D-$C_4$ is found more cohesive than 3D-$C_4$. This can be expected from the fact that graphitic-like carbon is more stable that 3D diamond-like Lonsdaleite 3D-$C_4$. Upon inserting extra carbon atoms (white spheres) leading to 3D-$C_6$ a decrease of the cohesive energy is observed due to expansion of the cell resulting into a rather open structure that will be reflected in the mechanical by a reduced hardness. At this point we mention that $C_6$ structure was used to model tricarbon $C_3$ molecule in the solid state [24]. The same stoichiometry $C_6$ with mixed $sp^3$-$sp^2$ in tetragonal system was identified with superhard and semi-conducting properties [25].

Alternative 2D-$C_4$ layered structure devised with eclipsed-like planes facing each other, i.e. oppositely to Fig. 1a, is shown in Fig. 2a. The space group is *P6cc,* No. 184, i.e., a lower symmetry than above studied allotropes, but the cohesive energy is close to 2D-$C_4$ above. The four carbon atoms in (4b) 1/3, 2/3, ¾ Wyckoff position are distributed with 2 C at z= ¼ and 2 C at z= ¾, leading to two horizontal planes. Inserting an extra carbon at z= ½ (white sphere) leads to $C_5$ stoichiometry with a configuration shown in Fig. 2b. Such a strained structure was geometry relaxed following the protocol described in Section 1. The resulting calculations led to the 3D structure shown in Fig. 2c. Crystal symmetry search led to assign the space group *P*-6*m*2 No. 187 with two 2-fold carbon positions and one Wyckoff position at the cell center. The crystal data are given in Table 1, last columns. The representation in Fig. 2d shows two cells along the *c* hexagonal axis exhibiting two successive *C3* (brown-white-brown carbons) linear entities aligned along *c*, oppositely to the *zigzag* arrangement in $C_6$.

With three different carbon sites, such a structure will be used as a template to examine the effect of different nitrogen substitutions for carbon leading to a two carbonitrides as shown below. Regarding the cohesive energies whilst it decreases with respect to 2D-$C_4$, the relevant result is the strong decrease of the volume of $C_5$ signaling condensation. Such result lets expect a dense carbon allotrope with large hardness, i.e., strong incompressibility. Lastly, the topological analysis using TopCryst program [2] showed **lon** topology, like Lonsdaleite as above mentioned $C_6$.

To summarize on the two carbon allotropes: Starting from staggered 2D planes on one hand, and from eclipsed-like 2D planes facing each other, on the other hand, three-dimensional 3D carbon allotropes were produced, and geometry optimized to their ground states, namely hexagonal $C_6$ and $C_5$ respectively. Energy-wise, while the 2D starting structures were found with slightly larger cohesive 2D-$C_4$ in space group *P6₃/mmc,* No. 194 (alike graphite) than 2D -$C_4$ in space group *P6mm,* No. 184, 3D $C_6$ was identified with much lower cohesion than $C_5$ (Table 1, last line). Crystallographically, the starting carbon planes different stacking of the 2D structures have been shown to have consequences on the 3D carbon structures involving vertically aligned *C3* motifs (eclipsed) in $C_5$ (Fig. 3a) versus *zigzag* -like (non-eclipsed or staggered) *C3* arrangement in $C_6$. Oppositely to the energy trends of the 2D pristine $C_4$ starting templates, 3D $C_5$ is found much largely cohesive than 3D-

$C_6$ as shown in Table 1. Their different stackings of *C3* units along the hexagonal direction will have drastic effects on the mechanical properties, specifically regarding the hardness.

### *2b- Binary phases*

$C_5$ was subsequently used as a template to devise binary C-N chemical systems by introducing nitrogen substitutionally. Namely, the following binary systems were obtained: $C_4N$ and $C_2N_2$. Table 2 presents the calculated lattice parameters, and the structures are shown in Figure 3 in ball-and-stick and polyhedral representations. These binary systems were obtained following a protocol that we now detail: The replacement of central carbon (white sphere in Fig. 2c) by nitrogen leads to carbon rich nitride $C_4N$ which shows a structure like pristine $C_5$ with small differences as for the slightly smaller volume and the lower cohesive energy. The shortest interatomic distance is observed for C-N with 1.35 Å but the C-C distances remain within range of $C_5$. Then, increasing the amounts of nitrogen leading to $C_3N_2$ and $C_2N_3$ stoichiometries led to little cohesive systems after full geometry relaxations. Also, dynamic instabilities (negative phonon acoustic modes, cf. section 4-2) were also observed. Therefore, $C_3N_2$ and $C_2N_3$ were discarded from further analyses. However, by removing body center nitrogen, i.e., at N(1*f*) position from $C_4N$ and letting C2 (2i) position be occupied by nitrogen thus leading to the $C_2N_2$ equiatomic nitride, a cohesive system (-1.66 eV/atom) was obtained with fully relaxed structure parameters given in the 2$^{nd}$ column of Table 2. The structure shown in Fig. 3c exhibits the peculiar feature of two puckered C-N layers forming corner sharing *C2N2* tetrahedra within the layer, as illustrated in the tetrahedral representation. The C-N distance is now larger than in $C_4N$, i.e., d(C-N) =1.46 Å and the interlayer distance is large as inferred from d(N-N) > 3 Å. Such features need further assessments that we approach with the charge densities and the electron localization projections in next Section 3.

3- **Projections of the charge densities and the electron localization: where are the electrons?**

*3-1 Charge density projections.*

For an illustration of the peculiar characteristics brought by nitrogen into the carbon lattice, the analysis is extended to a qualitative illustration of the charge densities. The difference of Pauling electronegativities ($\chi$): $\chi C = 2.55$ versus $\chi N = 3.04$ let expect charge transfer from C to N. The other relevant point is the hybridization characters of carbon in $C_5$ and $C_6$, i.e., $sp^3$ and $sp^2$. Figure 4 illustrates with yellow volumes the charge density. In Fig. 4a $C_6$ shows continuous charge density with characteristic $C(sp^3)$ and shifted *C3* vertical alignment with central $C(sp^2)$. Similar features of charge densities are observed in Fig. 4b for $C_5$ where the alignment of *C3* is not shifted but aligned with the nearest neighbor in the second cell as shown in the double cell of Fig. 3a RHS.

Significant charge density changes occur upon substituting central carbon for nitrogen, illustrated in Fig. 4c, with the large charge density localized on N, explained by its larger electronegativity versus C as mentioned above. This is also observed in Fig. 4d exhibiting the charge density on N pointing to the empty structure space along *c* hexagonal (vertical) direction. The labeling as carbonitrides of both $C_4N$ and $C_2N_2$ is justified chemistry-wise.

*3-2 Electron localization projections.*

We further elaborate on the electron distribution in the equiatomic $C_2N_2$ especially as regarding the role played by the nitrogen lone pair in the interlayer space. For this purpose, we used the scheme of the electron localization function (ELF) devised by Becke and Edgecomb [26] as initially proposed within Hartree–Fock calculations. It was later adapted to DFT methods by Savin et al. [27] as based on the kinetic energy in which the Pauli Exclusion Principle is included. ELF is defined as a normalized function with $0 \leq ELF \leq 1$. For $ELF = 0$ very low electron localization (blue zones) and $ELF = 1$ for strong localization (red zones corresponding to large grey volume. A free electron–gas like behavior corresponds to ELF ~½ with green zones (cf. ruler in Fig. 5). Figure 5 shows the corresponding 3D ELF of $C_2N_2$ with gray volumes on the left-hand side (LHS) and the 2D ELF slice on the right-hand side (RHS) with colors defined in the vertical ruler. The large gray volumes (LHS) of strong electron localization represent the nonbonding nitrogen lone pair (LP) pointing towards the interlayer space and less towards carbon where the C-N bonding occurs. RHS projection illustrates further the ELF with the large red area due to N(LP) covering almost all the empty space between the layers and leaving little zero localization (blue zones); it can be proposed that these features maintain the structure stability and let expect exceptional mechanical properties. The yellow areas qualifying medium localization is found between C and N, whence the bonding between them.

## 4- Physics of $C_5$ and the binary carbonitrides $C_4N$ and $C_2N_2$.

### 4-1 Mechanical properties from the elastic constants

An analysis of the mechanical behavior was carried out with the calculation of the elastic properties through inducing finite distortions of the lattice. The system is then fully described by the calculated sets of elastic constants $C_{ij}$ (i and j correspond to directions). Table 2 reports the elastic constants of $C_5$, the binary carbonitrides $C_4N$ and $C_2N_2$ as well as hexagonal $C_6$ presented in Section 2a (Table 1) for the sake of comparison. All $C_{ij}$ values are positive, signaling mechanically stable phases. $C_5$ has systematically larger $C_{ij}$ magnitudes versus $C_6$ that can be inferred from the two different structures involving vertically aligned *C3* motifs (eclipsed) in $C_5$ (Fig. 3a) versus *zigzag* (non-eclipsed) *C3* arrangement in $C_6$ (*vide supra*). The carbonitrides derived from $C_5$ also exhibit similarly large $C_{ij}$ magnitudes especially $C_4N$ whereas $C_2N_2$ shows smaller magnitudes. However, the large $C_{33}$ magnitude close to 500 GPa in such layered-like system can be highlighted and correlated to the repulsive action of the two nitrogen LP's facing each other, as discussed above with the ELF.

The $C_{ij}$ magnitudes are translated into mechanical properties trough deriving the bulk ($B_V$) and shear ($G_V$) moduli were obtained by Voigt's averaging [18] of the elastic constants using ELATE software [19]. The last two columns of Table 2 show the bulk and the shear moduli with values that follow the trends observed for $C_{ij}$. From the bulk and shear moduli, the ratio $G_V/B_V$ called the Pugh ratio [28] allows distinguishing ductile behavior ($G_V/B_V < 1$) from brittle behavior ($G_V/B_V > 1$). The Pugh ratios of the different chemical systems are given in Table 3 in the last but one column. With $G_V/B_V(C_6) = 0.87$ a ductile behavior can be assigned to $C_6$ due to its structural characteristics of *C3 zigzag* stacking along the hexagonal c axis. Oppositely $G_V/B_V(C_5) = 1.03$ larger than 1 and the derived nitrides with similar Pugh ratios can be considered as brittle. Using the Pugh ratios, we then calculated the Vickers hardness $H_V$ with Tian et al. model of microscopic theory of hardness [20]:

$$H_V = 0.92(G_V/B_V)^{1.137} \, G_V^{0.708}$$

The last column of Table 3 shows the calculated values of $H_V$. Expectedly, $H_V(C_6) = 49$ GPa is calculated much lower than $H_V(C_5) = 70$ GPa qualified consequently as super-hard but $H_V(C_4N) = 72$ GPa is slightly larger due to the smaller volume and the short C-N connections. Lastly, $H_V(C_2N_2) = 62$ GPa is a large magnitude despite the layered-like structure. It can be assumed that the high hardness arises from the strong bonding within the thick layers of *C2N2* tetrahedra.

### 4.2 - Dynamic properties from the phonons

A relevant criterion of phase dynamic stability is obtained from the properties of the phonons, defined as quanta of vibrations. Their energy is quantized with the Planck constant 'h' used in its reduced form, i.e., $\hbar$ with $\hbar = h/2\pi$, resulting into by $E = \hbar\omega$ where $\omega$ is the frequency. Figure 6 depicts in four panels the phonons band structures of the two allotropes $C_6$ and $C_5$ and the latter's derived nitrides $C_4N$ and $C_2N_2$. The bands develop along the major lines of the hexagonal Brillouin Zone (horizontal *x*-axis). The frequency along the y-axis is expressed in units of Terahertz (THz, with 1 THz = 33 cm$^{-1}$). There are 3N-3 optical modes found at higher energy than three acoustic modes that start from zero energy ($\omega = 0$) at the $\Gamma$ point (center of the Brillouin Zone), up to a few Terahertz. They correspond to the lattice rigid translation modes of the crystal (two transverse and one longitudinal). The remaining bands correspond to the optic modes. All phonon frequencies are found positive letting confirm the dynamic stability of the announced allotropes and the two nitrides. The highest frequencies in all panels are close to 40 THz, equivalent to 1650 cm$^{-1}$. This magnitude is exactly within the range of C=C vibrations with a strong signal, i.e., 1625–1680 cm$^{-1}$ in the molecular state as listed in the web-available "RAMAN Band Correlation Table". The second highest vibration is found with bands around 50 THz in $C_4N$, arising from the shorter C-N distance of 1.35 Å versus inter-carbon separation in $C_5$ with of 1.46 -1.54 Å (Table 1). Lastly, in Fig. 6c the flat band along $\Gamma$-A at E = 0 THz, i.e., along the vertical $k_z$ BZ direction of reciprocal space, is a signature of no dispersion of bands along the empty structural gap characterizing $C_2N_2$.

*4-3 Electronic band structures*

Figure 7 shows in four panels the band structures of the two allotropes and the two nitrides in the hexagonal Brillouin Zone. The zero energy along the y-axis is considered with respect to the Fermi level $E_F$ crossed by finite bands in $C_6$, $C_5$ and $C_4N$, i.e., with no separation between the valence band VB and the conduction band CB at $E_F$. These three systems are then considered as metallic as resulting from the presence of $C(sp^2)$ with dispersed π-like electrons.

An exception can be observed for the equiatomic $C_2N_2$ where direct band gap of ~5 eV between $A_{VB}$ and $A_{CB}$ and the top of the valence band VB. Here the energy reference $E_V$ is considered with respect to the top of VB. Such a result can be explained by the localization of charge density around N pointing through the LP towards the interlayer space.

**Author Contributions:** Conceptualization, methodology, investigation, formal analysis, visualization, writing the paper. No A.I. nor other web resources were used.

**Funding:** This research received no external funding.

**Data Availability Statement:** The data presented in this study are available upon reasonable request due to LGU restrictions, i.e., privacy, etc.

**Acknowledgments:** Computational facilities from the Lebanese German University are gratefully acknowledged.

**Conflicts of Interest:** The author declares no conflict of interest.

**TABLES**

Table 1 Crystal structure parameters of carbon allotropes.

| | 3D-C$_4$ $P6_3/mmc$ No. 194 | 2D-C$_4$ $P6_3/mmc$ No. 194 | 3D-C$_6$ $P6_3/mmc$ No. 194 | | 2D-C$_4$ $P6mm$ No. 184 | 3D C$_5$ $P$-$6m2$ No. 187 |
|---|---|---|---|---|---|---|
| $a$, Å | 2.505 | 2.460 | 2.484 | | 2.461 | 2.487 |
| $c$, Å | 4.169 | 6.732 | 6.953 | | 6.698 | 5.581 |
| $V_{cell}$, Å$^3$ | 22.66 | 35.29 | 37.14 | | 37.14 | 29.88 |
| *Shortest dist.* /Å | d(C-C) =1.54 | d(C-C) =1.42 | d(C1-C2) =1.45  d(C1-C1) =1.55 | | d(C-C) =1.42 | d(C-C') =1.46  d(C-C) =1.54 |
| Atomic position | C(4f) 1/3,2/3,0.06275 | C(4f)1/3,2/3,0.0 | C1(4f) 2/3,1/3,0.9579  C2(2d) 2/3,1/3, ¼ | | C(4b) 1/3,2/3, ¾ | C1 (2h) 1/3 2/3 0.861  C2 (2i) 2/3 1/3 0.762  C'(1f) 2/3 1/3 ½ |
| E(Total)eV  E$_{coh}$/at. eV | -36.28  -2.47 | -36.90  -2.63 | -48.88  -1.54 | | -36.72  -2.58 | -42.49  -1.89 |

Table 2. Crystal structure parameters of novel binary phases.

| $P\text{-}6m2$ No 187 | $C_4N$ | $C_2N_2$ |
|---|---|---|
| Topology | **lon** | 3,4**L**147 |
| $a$, Å | 2.521 | 2.377 |
| $c$, Å | 5.388 | 6.382 |
| $V_{cell}$, Å$^3$ | 29.66 | 31.25 |
| *Shortest d-d-* Å | d(C1-N)=1.35<br>d(C2-C2)=1.52<br>d(C1-C2)=1.57 | d(C-N)=1.46<br>d(C-C)=1.63<br>d(N-N)=3.73 |
| Atomic position | C1 (2h) 1/3 1/3 0.751<br>C2 1f) 2/3 1/3 ½ | C (2h) 1/3 2/3 0.872<br>N (2i) 2/3 1/3 0.792 |
| E(Total) eV<br>$E_{coh}$/at. eV | -41.99<br>-1.74 | -33.42<br>-1.66 |

Atomic energies (atoms in a large box): E(C ) = -6.6 eV, E (N) = -6.8 eV.

Table 3. Elastic constants $C_{ij}$, bulk $B_{Voigt}$, shear $G_{Voigt}$ modules, and Vickers hardness $H_{Vickers}$ are in GPa (Giga Pascals) unit.

| | $C_{11}$ | $C_{12}$ | $C_{13}$ | $C_{33}$ | $C_{44}$ | $C_{66}$ | $B_{Voigt}$ | $G_{Voigt}$ | $G_V/B_V$ | $H_{Vickers}$ |
|---|---|---|---|---|---|---|---|---|---|---|
| $C_6$ | 641 | 139 | 100 | 1525 | 251 | 306 | 387 | 337 | 0.87 | 49 |
| $C_5$ | 920 | 95 | 46 | 1453 | 412 | 333 | 407 | 423 | 1.04 | 70 |
| $C_4N$ | 851 | 108 | 159 | 1707 | 372 | 356 | 409 | 435 | 1.06 | 72 |
| $C_2N_2$ | 1056 | 140 | 52 | 498 | 458 | 268 | 344 | 356 | 1.03 | 62 |

**FIGURES**

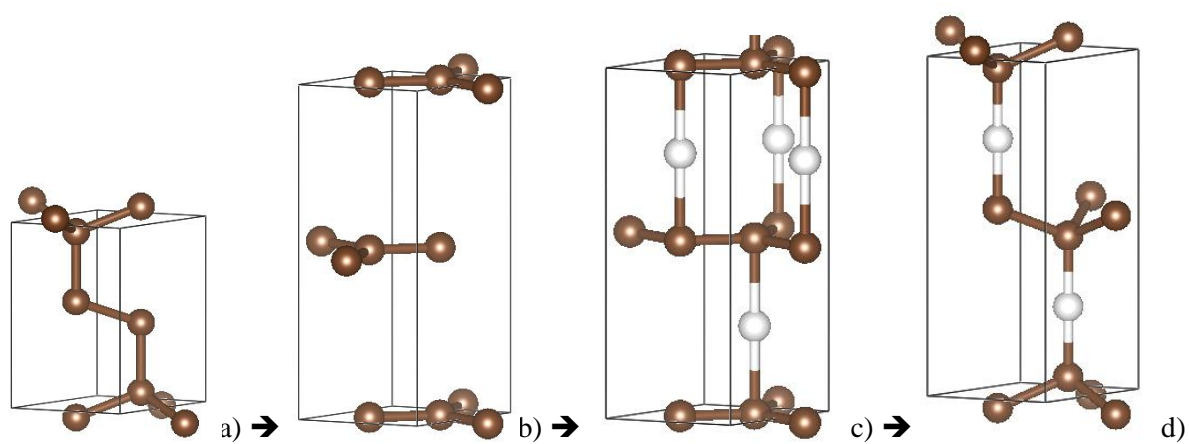

Figure 1. Schematic transformation from 3D C$_4$ Lonsdaleite (a) to layered 2D C$_4$ (b), then in (c) the subsequent insertion of carbon (white spheres), and d) the fully geometry optimized 3D C$_6$ structure. Space group $P6_3/mmc$, No. 194 (see text).

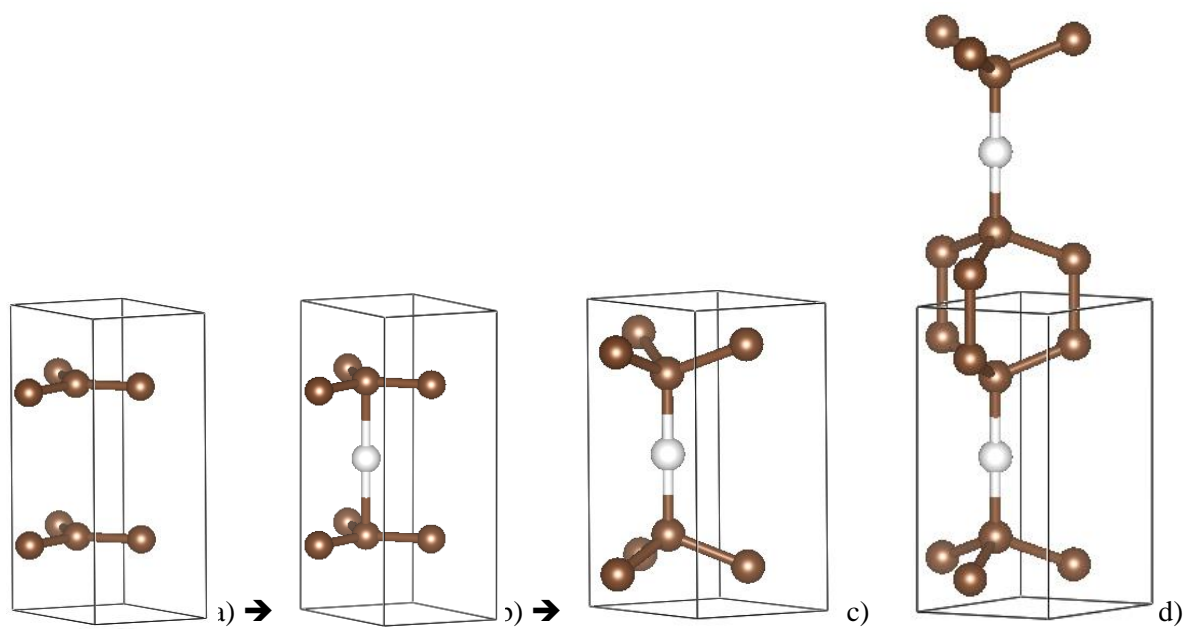

Figure 2: Schematic transformation from layered 2D C$_4$ (a) and the change of stoichiometry to C$_5$ (b) by inserting C at z= ½ (white sphere); c) the fully geometry optimized 3D C$_5$, with d) shwoing a double cell along c-hexagonal direction to highlight the alignment of *C3* (brown-white-brown spheres) motifs along the hexagonal vertical axis. Space group *P*-6*m*2, No. 187 (see text).

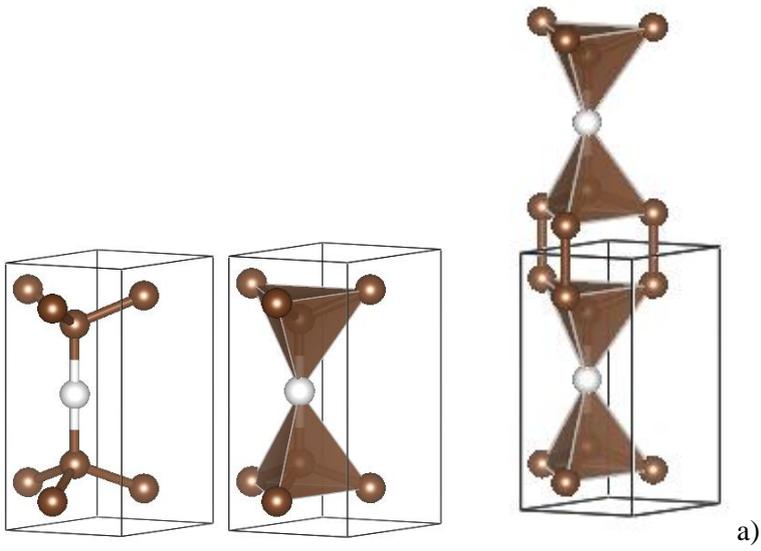

a)

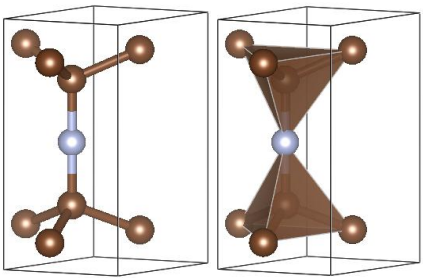

b)

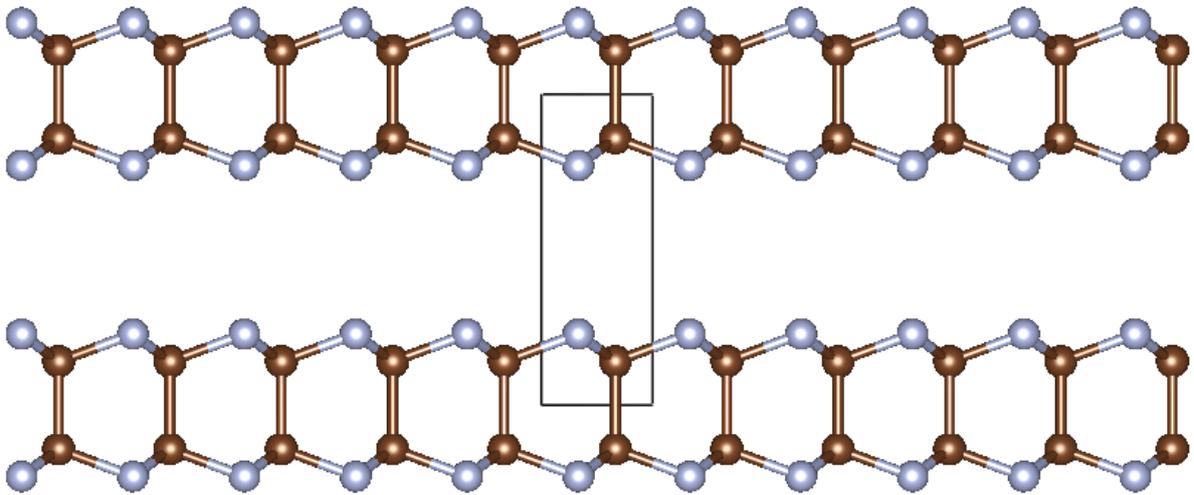

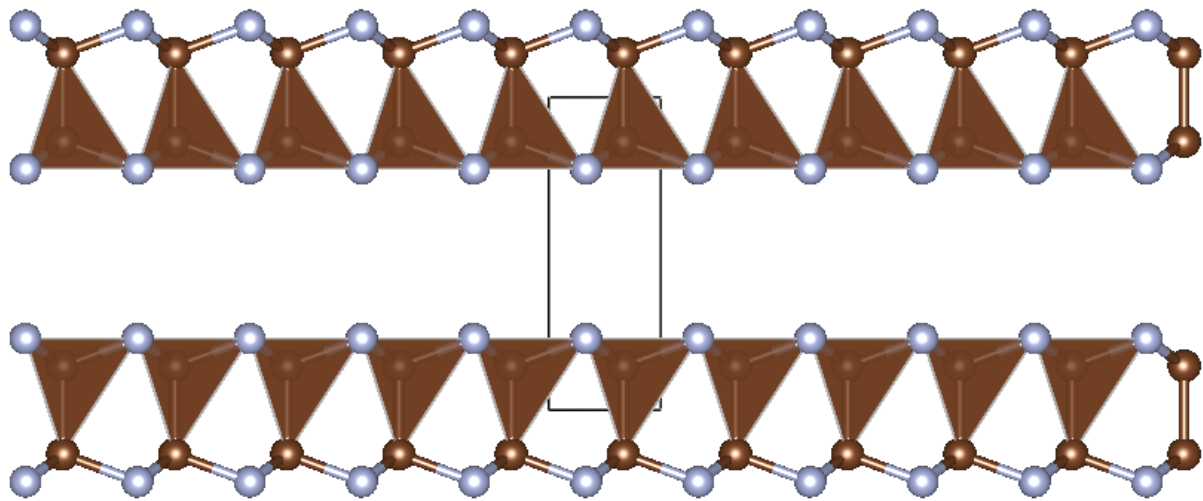

c)

Figure 3: Crystal structure in ball-and-stick and tetrahedral representations of a) $C_5$ (the white sphere illustrates the additional carbon as shown in Fig. 1), b) $C_4N$, c) $C_2N_2$. Brown, and grey spheres represent C and N respectively. Hexagonal c-axis is along the paper sheet.

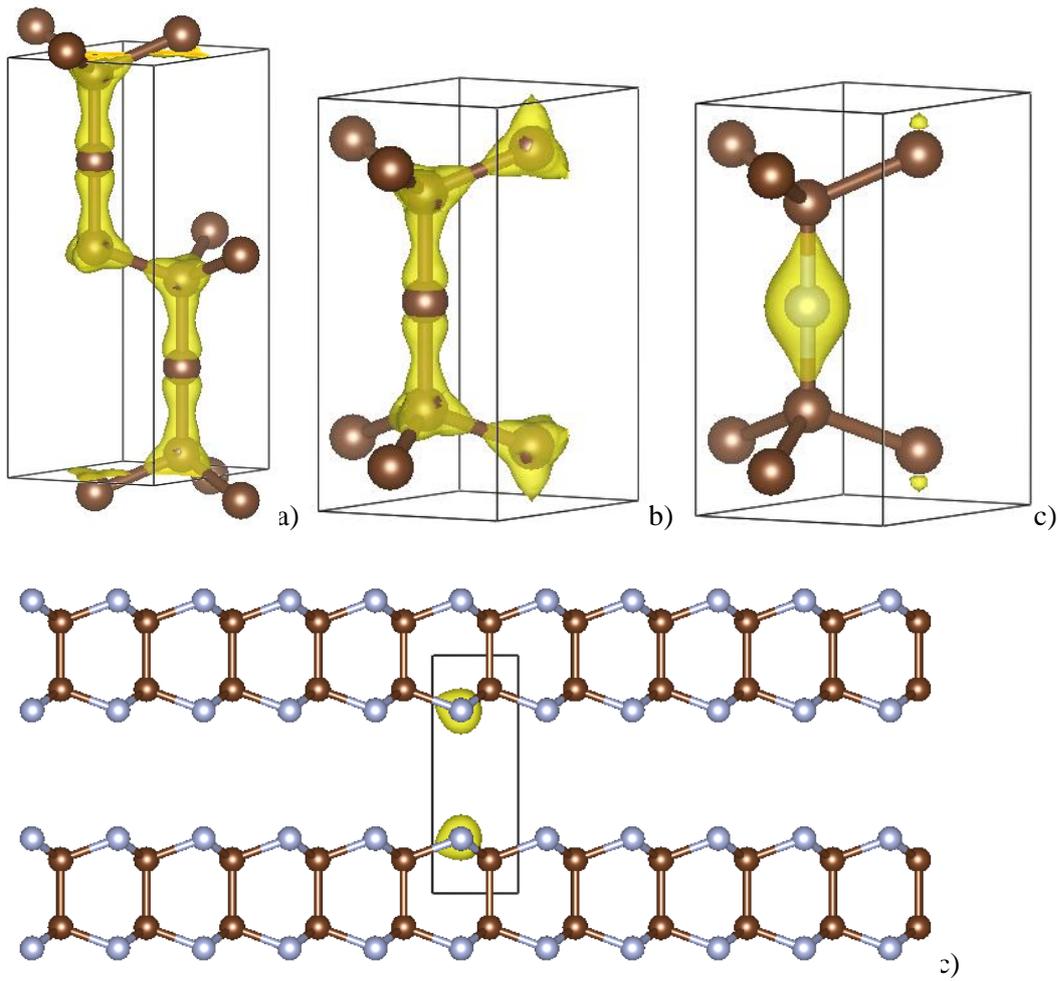

Figure 4: Charge density projections (yellow volumes) of a) $C_6$, b) $C_5$, b) $C_4N$, c) $C_2N_2$

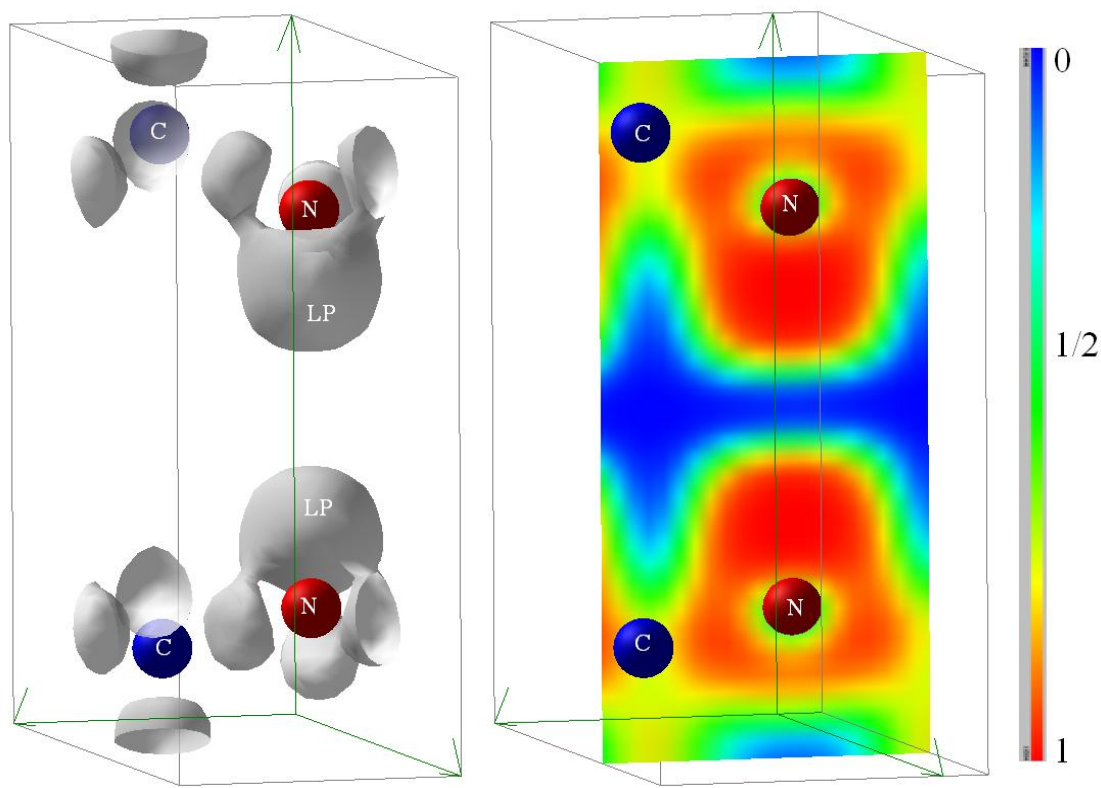

Figure 5. Electron Localization function ELF in equiatomic $C_2N_2$ nitride highlighting the large steric effect of non-bonding $N(2s^2)$-like lone pair (LP) on inducing the bilayer structure. ELF projections are shown in 3D grey volumes and 2D plane LHS (left hand side) and a plane slice along hexagonal direction RHS (right hand side). The ruler represents the ranges of ELF magnitudes from 0 to 1 with corresponding colors (cf. text).

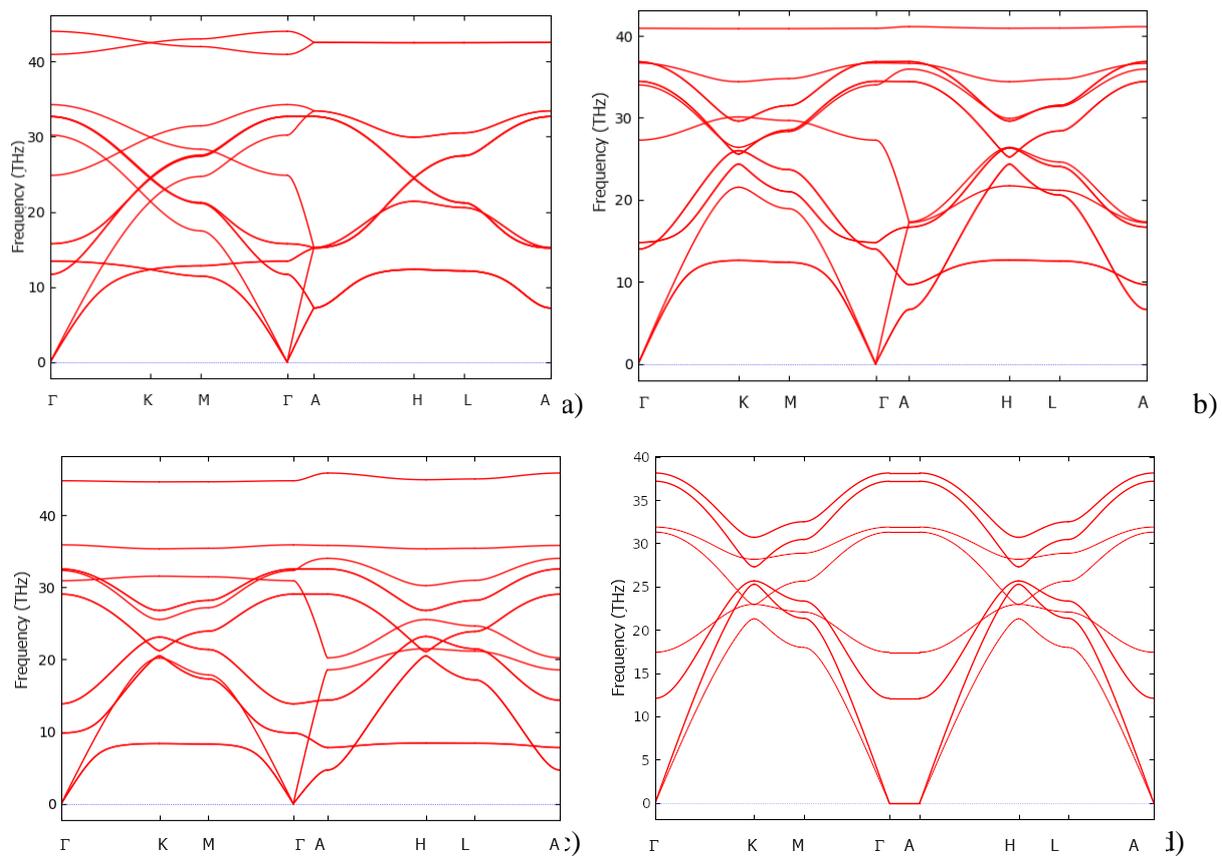

Figure 6: Phonons band structures of a) $C_6$, b) $C_5$, c) $C_4N$, d) $C_2N_2$

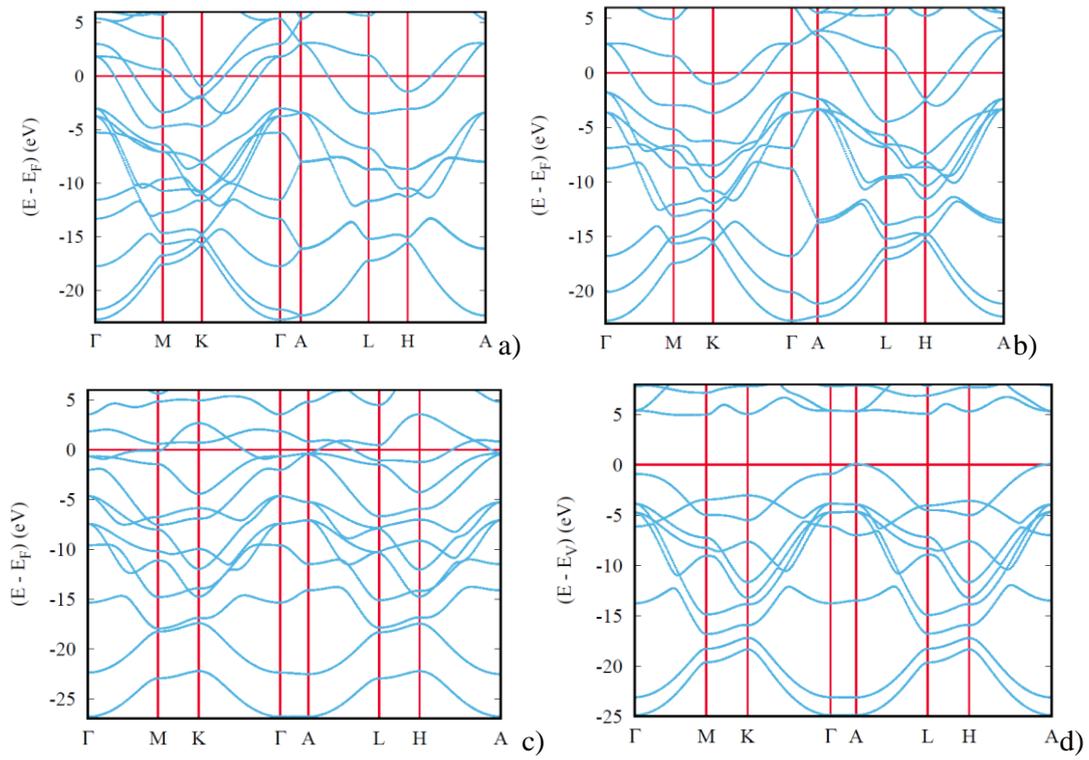

Figure 7: Phonons band structures of a) $C_6$, b) $C_5$, c) $C_4N$, d) $C_2N_2$